\documentclass[12pt]{article}
\usepackage{amsfonts}

\def\beq{\begin{equation}}
\def\eeq{\end{equation}}
\def\bea{\begin{eqnarray}}
\def\eea{\end{eqnarray}}
\def\ba{\begin{array}}
\def\ea{\end{array}}

\def\nn{\nonumber}

\def\ga{\gamma}

\def\si{\sigma}

\def\de{\delta}
\def\ka{\kappa}
\def\al{\alpha}

\def\la{\lambda}

\def\ve{\varepsilon}

\def\GeV{{\rm GeV}}

\def\ti{{\times\!}}

\def\cO{{\cal{O}}}
\def\cL{{\cal{L}}}

\newcommand{\ft}[2]{{\textstyle {\frac{#1}{#2}} }}

\begin{document}
\hfill AEI-2009-065\\

\centerline{\bf\Large Conformal invariance from non-conformal gravity}

\vspace{5mm}
\centerline{\bf Krzysztof A. Meissner${}^{1}$ and Hermann Nicolai${}^2$}

\vspace{5mm}
\begin{center}
{\it ${}^1$ Institute of Theoretical Physics, Faculty of Physics\\
University of Warsaw, Ho\.za 69, 00-681 Warsaw, Poland\\ ${}^2$
Max-Planck-Institut
f\"ur Gravitationsphysik (Albert-Einstein-Institut),\\
M\"uhlenberg 1, D-14476 Potsdam, Germany}
\end{center}

\begin{abstract}
\footnotesize{We discuss the conditions under which classically conformally
  invariant models in four dimensions can arise out of non-conformal
  (Einstein) gravity. As an `existence proof' that this is indeed
  possible we show how to derive $N=~4$ super Yang Mills theory with any compact
  gauge group $G$ from non-conformal {\em gauged} $N=4$ supergravity as a
  special flat space limit. We stress the role that the anticipated
  UV finiteness of the (so far unknown) underlying theory of quantum
  gravity would have to play in such a scheme, as well as the fact that
  the masses of elementary particles would have to arise via quantum
  gravitational effects which mimic the conformal anomalies
  of standard (flat space) UV divergent quantum field theory.}
\end{abstract}

\section{Introduction}

That conformal symmetry\footnote{Classic references are \cite{MS,CCJ};
  see also \cite{K} for a comprehensive review of conformal invariance
  in field theory.} might play a key role in resolving the hierarchy
problem remains a distinct possibility, especially in view of the
remarkable fact that the standard model (=SM) of elementary particle
physics \cite{N,P} {\em is} classically conformally invariant, except
for the explicit mass term in the scalar potential that is commonly
introduced for electroweak symmetry breaking. In such a `conformal
scenario' the observed mass scales of particle physics and their smallness
{\it vis-\`a-vis} the Planck scale might be explained solely via the
quantum mechanical breaking of conformal symmetry, in accordance with
the naturalness criterion of \cite{tH} (see also \cite{Bardeen}).
Simultaneously with conformal symmetry, the electroweak $SU(2)_w\times U(1)_Y$
symmetry of the SM would have to be broken via radiative corrections
by means of the Coleman-Weinberg (=CW) mechanism \cite{CW,Sher}.

While it is still not clear whether this idea can be made to work
in the realistic context of the SM, a main objection that has been
raised against it is that the SM couples to gravity and must eventually
merge into a theory of quantum gravity at the Planck scale.
Einstein's theory (with SM-like matter couplings) is certainly
{\em not} conformally invariant due to the presence of the dimensionful
coupling $\ka = M_P^{-1}$, and it is therefore far from evident how
a classically conformal Lagrangian might arise out of such a theory
at low energies. For this reason, recent and not so recent attempts
to incorporate scale invariance have proceeded from the assumption
that the `true' theory of gravity might be conformal (Weyl) invariant
gravity, that is, a theory invariant under {\em local} rescalings
of the metric and the matter fields, out of which Einstein's theory
might emerge only after spontaneous breaking of scale invariance
(see \cite{BD,BB} for an early proposal along these lines and for an
{\it entr\'ee} into the literature, and \cite{SZ,NR} for more recent work).
Such a theory would ultimately also be expected to involve terms
quadratic in the Weyl tensor. However, apart from the known difficulties
with $(Weyl)^2$ theories of gravity, the known ans\"atze at unification
in general do not give rise to effectively Weyl invariant low energy
theories\footnote{A possible exception is maximally extended conformal
  $N=4$ supergravity in four dimensions \cite{BdWdR}, but the status
  of this theory remains very unclear.}, despite the ubiquity of
dilaton-like fields in supergravity and superstring theory. For this
reason we here suggest a different route by exploring whether and under
what circumstances it may be possible to get a classically conformal
theory out of {\em non-conformal} Einstein gravity or some of its
supersymmetric extensions.

The specific example we focus on is the conformal theory {\em par
excellence}, maximally supersymmetric $N=4$ Yang-Mills theory in
four dimensions with compact gauge group $G$ \cite{BSS,GOS}. We will
demonstrate explicitly how this theory can be obtained as a
conformal limit of a non-conformal  theory, namely gauged $N=4$
supergravity \cite{BKS,dRW,SW} as $\ka\rightarrow 0$. Our
construction is inspired by a recent re-derivation from gauged
supergravities in three dimensions \cite{BdRHR,BHRS} of the
conformally invariant and globally supersymmetric ($N\leq 8$) models
thought to describe multiple M2 branes. To be sure, the present
construction only furnishes an `existence proof': the example of
$N=4$ super Yang Mills theory `overshoots' in that this theory is UV
finite to all orders \cite{finite1,finite2,West1}, hence exactly,
{\it i.e.} quantum mechanically conformal. Furthermore, the unbroken
supersymmetry entails that there can be no radiative symmetry
breaking \cite{FL,West}. Nevertheless, the derivation of this model
from a non-conformal theory of gravity does illustrate our main
point.

The present article is thus complementary to our recent proposal
\cite{MN1,MN4} to implement the CW mechanism in the SM~\footnote{There
  is a large literature on this subject, see e.g. \cite{Einhorn1} for
  references to earlier work.}. That work rests on two basic assumptions,
namely $(i)$ the absence of intermediate mass scales between the weak
scale and the Planck scale $M_P$; and $(ii)$ the requirement that the
RG evolved couplings exhibit neither Landau poles nor instabilities
over this whole range of energies. The first assumption (which is obviously
subject to experimental falsification) is necessary because any large
intermediate scale appearing in a quantum field theoretic context (such
as a GUT scale at $10^{16}\,\GeV$) is evidently at odds with classically
unbroken conformal invariance. The second hypothesis is to ensure the
`survival' of  the SM up to the Planck scale. As shown in \cite{MN1} this
requirement implies strong restrictions on the SM parameters. A further
important feature is that it may make electroweak symmetry breaking
mandatory due to the indirect coupling of the scalars to the strong
interactions \cite{MN4}.

The present ansatz based on classically unbroken conformal symmetry thus
pursues the same goal as low energy supersymmetry models, namely to explain
the emergence and stability of small scales in particle physics. In both
scenarios we must {\em assume} that the Planck scale theory of quantum
gravity is sufficiently benign so as not to affect low energy physics
in too drastic a manner. In supersymmetric models this is achieved in
part via the cancellation of quadratic divergences. However, in addition
to having to introduce a multitude of new (and so far unobserved) particles
and couplings one faces the notorious problem that supersymmetry  is
impossible to break spontaneously in a way that would fully conform
with low energy physics: in all `realistic'  scenarios, it must be
broken explicitly {\em by hand}, but there is no explanation why the soft
breaking terms are not Planck scale. By contrast, (classical) conformal
symmetry is not as averse to breaking by quantum effects, and allows
for greater economy in low energy model building. There, it is the
(postulated) structure of the anomalous Ward identity \cite{Bardeen}
(see eqn.~(\ref{Ward}) below) which ensures the absence of quadratic
divergences, hence of Planck scale mass terms and a Planck scale cosmological
constant in the low energy effective action. Consequently, mass terms and
symmetry breaking would arise solely from the logarithmic terms in the
effective potential induced by quantum corrections {\`a la}
Coleman--Weinberg. In this perspective the main issue is not only to
explain the embedding of the SM into an UV complete theory of quantum
gravity and quantum space-time, but also to understand how a UV finite
Planck scale theory can produce conformally anomalous corrections to
a classically conformal low energy effective action. Even though we have
so far no working model, we would thus expect that the CW mechanism in a
UV finite theory must undergo a metamorphosis and be replaced a gravitational
analog, in such a way that $\ka^{-1}$ acts as the effective cutoff. Conformal
symmetry would then be broken not because of the need to regulate UV
divergences, but because the quantum gravity theory into which the SM is
embedded is itself {\em not} conformal, but leaves its footprint in the
low energy effective action only in the form of logarithmic corrections.

This paper is organized as follows. In section~2 we review the basics of
gauged $N=4$ supergravity, and in section~3 we explain in detail how
to take the flat space limit of this theory in such a way that a classically
conformal theory emerges; these two sections contain our main technical
results. In the final section, we restate our main conjecture and discuss
possible avenues towards its solution.

\section{N=4 gauged supergravities}

In this section we review $N=4,d=4$ supergravities and their most general
gaugings \cite{BKS,dRW,SW}. For details we refer readers to \cite{BKS} whose
conventions and notations we follow almost without exception.~\footnote{We
  use $\al,\beta,\dots$ rather than $m,n,\dots$ for flat (Lorentz) indices
  in four dimensions.} The most general $N=4$  theory couples the
 gravitational $N=4$ multiplet
\beq\label{gravmult}
1\ti \left[2 \right] \,\oplus\, 4\ti\left[\ft32\right] \,\oplus\,
6\ti \left[1\right] \,\oplus\, 4\ti\left[\ft12\right] \,\oplus\,
2\ti \left[0\right]
\eeq
to $n$ vector multiplets of $N=4$ supersymmetry
\beq\label{vectmult}
n \times \bigg( 1\ti \left[1\right] \,\oplus\, 4\ti\left[\ft12\right] \,
  \oplus\,   6\ti \left[0\right] \bigg)
\eeq
where the spin is indicated in square brackets as $[s]$. The scalar
sectors describing the self-interactions of the two gravitational
scalars and the $6n$ scalars from the vector multiplets are governed
by non-linear $\si$-models over the coset spaces $SU(1,1)/U(1)$ and
$SO(6,n)/SO(6)\times SO(n)$, respectively. Here $SO(6)\cong SU(4)$
is part of the R symmetry $U(4)$ that rotates the four supercharges
\cite{Fer}, while $SO(n)$ acts as an outer automorphism that rotates
the $n$ vector multiplets. The gravitational coset $SU(1,1)/U(1)$ is
parametrized by one complex field $Z(x)$ and its complex conjugate
$Z^*(x)$ subject to the constraint $ZZ^*<1$, while the $6n$ matter
scalars coordinatize the second factor $SO(6,n)/SO(6)\times SO(n)$.
More concretely, the latter is described by a matrix $L_I{}^A(x)$
with curved indices $I,J,\dots$ and flat indices $A,B,\dots$
assuming the values $1,\dots,6+n$. This matrix transforms in the
usual way as
\beq
L(x) \rightarrow g L(x) h(x)\qquad
\mbox{with $g\in SO(6,n)$ and $h(x)\in SO(6)\times SO(n)$}
\eeq
under rigid $SO(6,n)$ and under local $SO(6)\times SO(n)$.
For the further analysis we need to split the (flat) matrix indices as
\beq
L_I{}^A = \big(L_I{}^{ij}, L_I{}^a \big)\qquad i,j=1,...,4 \;,\;\;
   a=1,\dots , n
\eeq
with the $SO(n)$ indices $a,b,\dots$. For the first six components,
we have exploited the local isomorphism $SO(6)\cong SU(4)$ to replace
the $SO(6)$ vector indices by an antisymmetric pair $[ij]$ of $SU(4)$
indices with the self-duality constraint
\beq
L_I{}^{ij} = (L_{I\,ij})^* = \frac12 \ve^{ijkl} L_{I\,kl}
\eeq
(We are here using the standard convention that complex conjugation
of the $SU(4)$ tensors corresponds to raising or lowering indices,
while the position of the $SO(n)$ indices does not matter.) The
inverse matrix $L^I{}_A$ then obeys $L_I{}^A L_A{}^J = \de_I^J$, or
\beq
- L_I{}^{ij} L_{J\,ij} + L_I{}^a L_{J\,a} =  \eta_{IJ}\;\;,\quad
L_I{}^{ij} L^I{}_{kl} = \de^{ij}_{kl} \;\;,\quad
L_I{}^a L^I{}_{ij} = 0
\eeq
where
\beq
\eta_{IJ} = {\rm diag}\, \big( -1,-1,-1,-1,-1,-1\,;\, +1, \cdots, +1 \big)
\eeq
is the Cartan-Killing metric on $SO(6,n)$.
The bosonic fields of the theory thus consist of the vierbein
$e_\mu{}^\al$ (with the metric $g_{\mu\nu} = e_\mu{}^\al e_{\nu\al}$),
the $(6+n)$ vector fields $A_\mu{}^I$ and the scalars $(Z,Z^*)$
and $L_I{}^A$. The fermionic sector contains four gravitinos
$\psi_\mu^i$, four `dilatinos' $\chi^i$ and $4n$ spin-$\frac12$
matter fermions $\la^{ai}$; these fermionic fields are subject to
\beq
\ga^5 \psi_\mu^i = + \psi_\mu^i \;\; , \quad
\ga^5 \chi^i = - \chi^i \;\; , \quad \ga^5 \la^{ai} = + \la^{ai}
\eeq
With the gravitational coupling (=inverse Planck mass) $\ka$ of
(length) dimension $cm$, the canonical dimensions of these fields
and the supersymmetry parameters $\ve^i$ are as follows:
\bea
[g_{\mu\nu}] &=&[e_\mu{}^\al] =  [L_I{}^A] = [Z] = 0 \;\; ,
\quad [A_\mu{}^I] = -1 \nn\\
{} [ \psi_\mu^i] &=& [\chi^i] = [\la^{ai}] = -3/2 \;\; , \quad
[\ve^i] = + 1/2
\eea
In passing we note that the torus reduction of pure half maximal $D=10$
supergravity from ten to four dimensions would give rise to a theory with
six vector multiplets and thus 12 vectors $A_\mu{}^I$, with the first
six vectors corresponding to the Kaluza-Klein vectors arising from
the $D=10$ metric $G_{MN}$, and the remaining six from the 2-form
field $B_{MN}$. The resulting theory would thus yield the coset
space $SO(6,6)/SO(6)\times SO(6)$.

The Lagrangian of the theory is obtained in the usual way by means of the
Noether procedure \cite{BKS}. We will here give it right away for the
{\em gauged} theory. This means that one promotes a subgroup of the global
(rigid) symmetry group $SO(6,n)$ to a local group using the vector fields
$A_\mu{}^I$ as Yang-Mills fields. In order to preserve the local $N=4$
supersymmetry of the original (ungauged) Lagrangian certain consistency
conditions must be obeyed.

Let us first present the Maurer-Cartan forms. For the coset $SU(1,1)/U(1)$
they are given by
\beq
Q_\mu = -\frac12 \big( \Phi\partial_\mu\Psi^* + \Psi\partial_\mu\Phi^* \big)
\;\; , \quad
P_\mu = \frac12 \big( \Phi\partial_\mu\Psi - \Psi\partial_\mu\Phi\big)
\eeq
with
\beq\label{Z}
\Phi := \frac{1-Z^*}{\sqrt{1- ZZ^*}} \;\; , \quad
\Psi := \frac{1+ Z^*}{\sqrt{1-ZZ^*}}
\eeq
Here the (imaginary) vector $Q_\mu$ is the $U(1)$ connection while
the vector $P_\mu$ corresponding to  the two coset degrees
of freedom is complex. For the matter coset $SO(6,n)/SO(6)\times SO(n)$
the Maurer-Cartan forms are
\bea\label{MC}
P_{\mu\,a}{}^{ij} &=& L^I{}_a \Big(\partial_\mu \de_I^K +
     f_{IJ}{}^K A_\mu^J\Big) L_K{}^{ij} \nn\\
Q_{\mu\,ab} &=& L^I{}_a \Big(\partial_\mu \de_I^K +
     f_{IJ}{}^K A_\mu^J\Big) L_{K\,b} \nn\\
Q_{\mu\,j}{}^{i} &=& L^{I\,ik} \Big(\partial_\mu \de_I^K +
     f_{IJ}{}^K A_\mu^J\Big) L_{K\, kj}
\eea
Here we have already included the gauge couplings via the structure
constants $f_{IJK}$ of the gauge group (we absorb the gauge coupling
constants into $f_{IJK}$). The integrability (Maurer-Cartan) relations
that follow from this definition are given in eq.~(10) of \cite{BKS}.
The structure constants must be completely antisymmetric after lowering
the index $K$ \cite{BKS}
\beq
f_{IJ}{}^L \eta_{KL} = f_{[IJ}{}^L \eta_{K]L}
\eeq
The quantities $(Q_{\mu\,ab}, Q_{\mu j}{}^i)$ are the `composite' gauge
connections for the local $SO(6)\times SO(n)$. In the ungauged theory
($f_{IJK}=0$) the fermions couple to the scalar fields {\em only} via
the $Q$'s and $P$'s. The full covariant derivatives are
\bea
D_\mu \psi_\nu^i &= & \partial_\mu \psi_\nu^i +
    \frac14 \omega_\mu{}^{\al\beta} \ga_{\al\beta} \psi_\nu^i
    - \frac12 Q_\mu \psi^i_\nu \nn\\
D_\mu \chi^i &=& \partial_\mu \chi^i +
   \frac14 \omega_\mu{}^{\al\beta} \ga_{\al\beta} \chi^i
    + Q_{\mu\j}{}^i \chi^j  + \frac32 Q_\mu \chi^i \nn\\
D_\mu \la^i_a &=& \partial_\mu \la^i_a + \frac14 \omega_\mu{}^{\al\beta}
   \ga_{\al\beta} \la^i_a + Q_{\mu\, a}{}^b \la^i_b +
   Q_{\mu\,j}{}^i \la^j_a + \frac12 Q_\mu \la^i_a
\eea
with the usual spin connection $\omega_\mu{}^{\al\beta}(e)$.
The quantities $P_\mu$ belonging to the coset, on the other hand,
appear in the kinetic terms of the scalar fields and in the Noether
couplings to the fermions.

While the ungauged theory has only derivative couplings to the scalar
fields, in the gauged theory there appear Yukawa-like couplings  of
the fermions as well as a (non-linear) potential for the scalar fields.
These new couplings (which vanish when the gauge couplings are set to zero)
involve the dimensionless quantities
\bea\label{C}
C_{ij} &:=& f_{IJK}  L^I{}_{ik} L^J{}_{jl} L^{K\,kl} = C_{ji} \nn\\
C_{aj}{}^i &:=& f_{IJK} L^{I\, ik} L^J{}_{kj} L_{K\,a} \nn\\
C_{ab}{}^{ij} &:=& f_{IJK} L^I{}_a L^J{}_b L^{K\,ij} = - C_{ba}{}^{ij}
\eea
These are just a variant of the so-called `T tensor' introduced
in \cite{dWN}. The preservation of local supersymmetry then requires
that the following identities must be satisfied
\bea
C^{ai}{}_k C_{aj}{}^k - \frac49 C^{ik} C_{kj} - {\rm trace} &=& 0 \nn\\
C^{ab}{}_{k(i} C_{bj)}{}^k - \frac23 C^{ak}{}_{(i} C_{j)k} &=& 0
\eea
In addition, these tensors must satisfy the differential identities
given in eq.~(12) of \cite{BKS}. Together these identities restrict
the possible choices of consistent gauge groups $\subset SO(6,n)$.
While only specific examples were studied in \cite{BKS,dRW}, more
recent work \cite{SW} based on embedding tensor techniques (see
\cite{dWNS} and references therein) has led to a more systematic
classification of gauge groups (which can be compact, non-compact
or non-semisimple). As a special case one also recovers the theories
obtained by torus reduction of $D=10$ supergravity coupled to $k$
vector multiplets with a Yang Mills gauge group $G$ of dimension
$= k$. This would give global $SO(6,6+k)$ in four dimensions
such that $G\subset U(1)^6\times SO(k)$.

Modulo terms quartic in the fermions the full Lagrangian has been
derived in \cite{BKS,dRW} and we here simply quote the result from
\cite{BKS}. Because we are here interested in the flat space limit
we re-instate the dimensionful coupling $\ka$ in all formulas.
We split the Lagrangian as
\beq
\cL = \cL_0 + \cL_{gauge}
\eeq
with
\bea\label{L0}
e^{-1}\cL_0  &=& \frac1{2\ka^2} R -
    \bar{\psi}_\mu^i \ga^{\mu\nu\rho} D_\nu \psi_{\rho i}
   - \frac12 a_{IJ} F_{\mu\nu}^{+}{}^I  F_{\mu\nu}^{+}{}^J \nn\\
          && -\, \frac12 \bar{\chi}^i \ga^\mu D_\mu \chi_i -
        \frac1{\ka^2 }P_\mu^* P^\mu
       -\frac12 \bar{\la}^{ai} \ga^\mu D_\mu \la_{ai}
       - \frac1{\ka^2}P_\mu^{a ij} P^\mu_{a ij} \nn\\
     && + \, \bar{\chi}^i\ga^\mu\ga^\nu \psi_{\mu i} P^*_\nu
       - 2i \bar{\la}^{ai} \ga^\mu\ga^\nu \psi_\mu^j P_{\nu \, aij} \nn\\
    && + \, \ka \Phi^{-1} F_{\mu\nu}^+{}^I
       \Big[ \frac{i}2 \bar{\psi}^\la_i \ga_{[\la} \ga^{\mu\nu} \ga_{\tau]}
       \psi_j^\tau L_I{}^{ij} -
\frac{i}2 \bar{\psi}^{\la i} \ga^{\mu\nu} \ga_\la \chi^j L_{I\,ij} \nn\\
&& \quad +\, \frac12 \bar{\psi}_{\la i} \ga^{\mu\nu} \ga^\la \la^{ai} L_{Ia}
     + \frac12 \bar{\la}^a_i \ga^{\mu\nu} \chi^i L_I{}^a
     + \frac{i}2 \bar{\la}^{ai} \ga^{\mu\nu} \la^{aj} L_{I\,ij} \Big]
 \nn\\
    && + \; h.c.
\eea
with
\bea
F_{\mu\nu}{}^I &:=& \partial_\mu A_\nu{}^I - \partial_\nu A_\mu{}^I
       + f^I{}_{JK} A_\mu{}^J A_\nu{}^K \nn\\
F^\pm_{\mu\nu} &:=& \frac12 \left( F_{\mu\nu} \pm \frac{i}2 \ve_{\mu\nu\rho\si}
                                       F^{\rho\si}\right)
\eea
and
\beq
a_{IJ} := \frac{1+Z}{1-Z^*} L_I{}^{ij} L_{J\, ij}
          + \frac{1+Z^*}{1-Z} L_I{}^{a} L_{J\, a}
\eeq
All terms proportional to the gauge couplings are assembled into
$\cL_{gauge}$ (and thus absent in the ungauged theory), {\it viz.}
\bea\label{Lg}
e^{-1} \cL_{gauge} &=& -\frac{2i}{3\ka} \Phi C_{ij}
   \left( \bar{\psi}^i_\mu \ga^{\mu\nu}
   \psi_\nu^j - \bar{\psi}_\mu^i\ga^\mu \chi^j -
   \bar{\la}^{ai} \la^j_a \right) \nn\\
 && +\,\frac{2i}{\ka} \Phi C_{aj}{}^i
     \left( \bar{\psi}^j_\mu\ga^\mu \la^a_i +
     \bar{\chi}^j \la^a_i \right)
    - \frac{2i}{\ka} \Phi C_{ab}{}^{ij} \bar{\la}^a_i \la^b_j \nn\\
 && -\frac1{2\ka^4}  \Phi^*\Phi \left( C_{aj}{}^i C^{aj}{}_i -
      \frac49 C^{ij} C_{ij} \right) + h.c.
\eea
The last term in (\ref{Lg}) is the {\em scalar potential}: as is obvious
from the expression given it is unbounded from below --- a standard
feature of gauged supergravity potentials.

The local supersymmetry variations are given by (modulo cubic terms)
\bea\label{SV}
\de e_\mu{}^\al &=& \ka \bar\ve^i \ga^\al\psi_{\mu i} + h.c. \nn\\
\de \psi_\mu^i  &=& \frac1{\ka} D_\mu \ve^i +
\frac{i}{2\Phi} \ga^{\mu\nu} F_{\mu\nu}{}^I L_I{}^{ij} \ve_j
     + \frac{2i}{3\ka^2} \Phi^* \ga_\mu C^{ij} \ve_j \nn\\
\de A_\mu^I &=& - 2i L^I{}_{ij} \left( \bar{\ve}^i \psi_\mu^j -
    \frac12 \bar{\ve}^i \ga_\mu \chi^j\right)
    + \Phi L^I{}_a \bar{\ve}^i \ga_\mu \la^a_i  + h.c. \nn\\
\de L^{Ia} &=& 2i\ka L^I{}_{ij} \bar{\ve}^i \la^{aj} + h.c. \nn\\
\de L^I{}_{ij} &=& -2i \ka \bar{\ve}_{[i} \la^a_{j]} L^I{}_a -
   {\rm dual} \nn\\
\de \chi^i &=& \frac{i}{\Phi^*} \ga^{\mu\nu} F_{\mu\nu}{}^I L_I{}^{ij} \ve_j
    + \frac2{\ka} \ga^\mu P_\mu \ve^i + \frac4{3\ka^2} \Phi C^{ij} \ve_j \nn\\
\de \la^{ai} &=& -\frac1{2\Phi^*} \ga^{\mu\nu} F_{\mu\nu}{}^I L_I{}^a \ve^i
    + \frac{2i}{\ka} \ga^\mu P_\mu{}^{a\,ij} \ve_j
    -\frac{2i}{\ka^2} \Phi C^{ai}{}_j \ve^j \nn\\
\de \Phi &=& - \ka\Phi^* \bar{\ve}_i \chi^i
\eea

\section{The conformal limit}

For the flat space limit we take $\ka\rightarrow 0$.
With the usual metric ansatz
\beq\label{gmunu}
g_{\mu\nu} = \eta_{\mu\nu} + \ka h_{\mu\nu}
\eeq
the curved space-time is flattened out in this limit, and gravity
decouples from the matter fields. Next, we must make sure that not
only gravity, but the full gravitational supermultiplet decouples
from the $n$ vector multiplets in this limit. The gravitino variation
contains the derivative on $\ve^i$, which comes with a factor of $\ka^{-1}$;
to keep this finite in the limit we must demand
\beq\label{De}
D_\mu \ve^i = 0 \;\; \Rightarrow\quad \ve^i(x) = const
\eeq
As expected, the limiting theory, if it exists, can only be
globally (rigidly) supersymmetric. For the scalar fields, we have
a formula analogous to (\ref{gmunu}),
 \beq
Z = \ka z \;\; , \quad
L_I{}^A = \left(\exp \big(\ka \phi \big)\right)_I{}^A =
\de_I^A + \ka \phi_I{}^A + \dots
\eeq
where the redefined scalar fields $z$ and $\phi_I{}^A$ are now of
dimension $(cm)^{-1}$ (like the vector fields) and the range of $z$
becomes $|z|<\ka^{-1}$. In the limit $\ka\rightarrow 0$ both cosets
are thus flattened; in particular, from (\ref{Z}) we see that the
fields $Z(x)$ decouple and $\Phi,\Psi\rightarrow 1$. Furthermore,
\beq
SO(6,n)/SO(6)\times SO(n) \longrightarrow {\mathbb{R}}^{6n}
\eeq
In the triangular gauge (where the `diagonal' components of $\phi_I{}^A$
vanish), the $6n$ scalar fields $\phi_a{}^{ij}$ from the $n$ vector
multiplets thus become coordinates of ${\mathbb{R}}^{6n}$. For the
matrix $L_I{}^A$ we obtain
\beq\label{Lka}
L_{ij}{}^{kl} = \de_{ij}^{kl} + \cO(\ka^2) \;\; , \quad
L_{{a}}{}^b = \de_a^b + \cO(\ka^2) \;\; , \quad
L_a{}^{ij} = \ka \phi_a{}^{ij} + \cO(\ka^3)
\eeq
(for notational simplicity, we do not distinguish anymore between flat
and curved indices here).

To obtain a {\em conformal theory}, however, (\ref{gmunu}) is not enough.
To see that extra requirements are needed we note that non-compact and
non-semisimple gaugings (the latter were not considered in \cite{BKS,dRW},
but see \cite{SW}) cannot yield a conformal limit: both of these gaugings
would involve Lie algebra generators in the non-compact part of $SO(6,n)$,
and thus mix the six gravitational with the $n$ matter multiplet vectors
in such a way that the gravitational vectors cannot decouple. This can
be directly seen by expanding the Maurer-Cartan connections, cf. (\ref{MC}),
making use of (\ref{Lka})
\bea
Q_{\mu\,ab} &=& f_{abc} A_\mu{}^c + f_{ab\,ij} A_\mu{}^{ij} + \cO(\ka^2)
     \nn\\
Q_{\mu\,j}{}^i &=& f_{jk \; lm}{}^{ik} A_\mu{}^{lm} +
             f_{jk\, c}{}^{ik} A_\mu{}^c + \cO(\ka^2)
\eea
In order to decouple the six vectors from the gravitational multiplet
(\ref{gravmult}) we must therefore demand that all components of $f_{IJK}$
vanish except for $f_{abc}$. In other words the gauge group must be
entirely contained in the $SO(n)$ symmetry group that rotates the $n$
vector multiplets (\ref{vectmult}) and commutes with the local $N=4$
supersymmetry; that is, we have~\footnote{An analogous restriction
  is found in the construction of \cite{BdRHR,BHRS}, and explains
  why only free field theories (which are trivially conformal)
  can arise when this limit is applied to $D=3$ gauged supergravities with
  $N>8$, that is, why globally supersymmetric interacting conformal
  theories in three dimensions exist only for $N\leq 8$. Here it follows
  similarly that no non-trivial flat space limit with residual rigid
  supersymmetry exists for $N>4$ supergravities in four dimensions.}
\beq\label{GSO(n)}
G\subset SO(n) \subset SO(6,n) \quad , \qquad {\rm dim}\, G = n
\eeq
Let us mention that the exclusion of non-compact and non-semisimple gauge
groups also ensures the positive definiteness of the kinetic terms of the
vector fields (whereas otherwise the flat space limit would suffer from
indefinite kinetic terms, unlike the original supergravity theory where
non-compact or non-semisimple gauge groups are compatible with positive
kinetic terms for the gauge fields). According to \cite{BKS} one possible
set of consistent gaugings is obtained with gauge groups $SU(2)\times SU(2)
\times G$ where $SU(2)\times SU(2)\subset SO(6)$, and $G$ is of dimension $n$;
the relevant gaugings here are thus obtained by setting the $SU(2)$
gauge couplings equal to zero. For the theories obtained by dimensional
reduction of $D=10$ type I supergravity coupled to $k$ vector multiplets
to four dimensions we have $G= U(1)^6\times H \subset SO(6+k)$ where
$H$ is the (compact) gauge group in ten dimensions.

Let us now return to the Lagrangian and supersymmetry variations.
First, we note that with (\ref{GSO(n)}), $L_I{}^A = \de_I{}^A$
(where $\phi_a{}^{ij} = 0$) is a fully supersymmetric stationary point
with vanishing cosmological constant. Namely, from (\ref{C}) it follows
immediately that all $C$-tensors vanish at this point, hence both
the extremization and the supersymmetry condition eqs.~(20) and (21) of
\cite{BKS} are trivially satisfied (this would not be so for other
choices of $G$). Let us then discuss the terms that decouple `trivially',
namely all those terms in (\ref{L0}) and (\ref{SV}) that do not involve
the gauge couplings, and hence are independent of the choice of gauge
group. With (\ref{gmunu}) the vierbein variation obviously reduces to
that of a free spin-2 field. In the gravitino variation, the first term
is absent due to (\ref{De}) while the second term starts at $\cO(1)$.
Splitting the index $I$, we see that the field strengths $F_{\mu\nu}{}^{ij}$
of $\cO(1)$ belong to the gravitational multiplet. The coupling to the
matter vectors $A_\mu{}^a$, on the other hand, occurs via the matrix
element $L_a{}^{ij}$; by (\ref{Lka}) it is therefore suppressed by
an extra factor of $\ka$, hence vanishes in the limit, as required.
The same mechanism is at work in the variation $\de A_\mu{}^I$: the
Kaluza Klein vectors pair up with $\big(\psi_\mu^i,\chi^i\big)$ while
the $6n$ matter vector terms have at least one extra factor of $\ka$.
The remaining variations are dealt with similarly. Idem for the
gauge-independent terms collected in the Lagrangian (\ref{L0}); for
instance, with (\ref{GSO(n)}) the coset part of the Mauer Cartan form becomes
\beq
P_{\mu\,a}{}^{ij} = \ka{\cal D}_\mu \phi_a{}^{ij} + \cO(\ka^3) \equiv
\ka\left( \partial_\mu \phi_a{}^{ij} +
             f_{abc} A_\mu{}^b \phi^{c\,ij}\right)  + \cO(\ka^3)
\eeq
hence
\beq
\frac1{\ka^2} P_\mu^{a\,ij} P^\mu_{a\,ij} =
{\cal D}_\mu \phi_a{}^{ij} {\cal D}^\mu \phi^a{}_{ij} + \cO(\ka^2)
\eeq
and the scalar kinetic term acquires the requisite form for
$\ka\rightarrow 0$. Again it is straightforward to see that all the
other matter terms with derivative scalar couplings decouple from the
supergravitational degrees of freedom in this limit.

The `non-trivial' terms are the ones involving the gauge couplings and the
non-derivative scalar couplings. In order to analyze the relevant
contributions we expand the $C$-tensors of (\ref{C}) in terms of the $6n$
scalars $\phi_a{}^{ij}$ in (\ref{Lka}). From inspection of the scalar
potential and the Yukawa terms in (\ref{Lg}) we see that the potentially
dangerous contributions are the ones involving inverse powers of $\ka$,
and we must therefore ensure that these terms are absent before taking
$\ka\rightarrow 0$. With the choice of gauge group (\ref{GSO(n)}) we find
\bea
C_{ij} &=& \ka^3 f_{abc} \phi_{a\,ik} \phi_{b\, jl} \phi_c{}^{kl}
  + \cO(\ka^4) \nn\\
C_{aj}{}^i &=& \ka^2 f_{abc} \phi_b{}^{ik} \phi_{c\,jk} + \cO(\ka^3)  \nn\\
C_{ab}{}^{ij} &=& \ka f_{abc} \phi_c{}^{ij} + \cO(\ka^2)
\eea
Again we see that the choice of gauge group is crucial in order to
ensure consistency of the decoupling limit: if there was a
non-vanishing structure constant $f_{ab\,ij}$ mixing the $SO(6)$ and
$SO(n)$ subgroups $C_{ij}$ would start at $\cO(\ka^2)$ instead, and for
instance the scalars would not decouple in the gravitino variation and
other terms. We next substitute these expansions into the Lagrangian
and the supersymmetry variations. For the potential we get
\beq\label{Pot}
 \frac1{2\ka^4}  \Phi^*\Phi \left( C_{aj}{}^i C^{aj}{}_i -
      \frac49 C^{ij} C_{ij} \right) =
\frac12 f_{abe} f_{cde} \phi^{a\,ik} \phi^b{}_{jk} \phi^{c\,jl} \phi^d{}_{il}
+ \cO(\ka)
\eeq
To see that this is just the usual potential of $N=4$ Yang Mills
theory in terms of six real Lie algebra valued scalar fields $X^{m\,a}$
(with $m,n=1,\dots,6$) we make use of the six real antisymmetric
matrices $(\al^r_{ij},\beta^r_{ij})$ for $r=1,2,3$ (explicit expressions
are given, for instance, in \cite{BSS,GOS})
\beq
\phi^a{}_{ij} = \sum_{r=1,2,3} \left( \al^r_{ij} X^{r\,a} +
               i\beta^r_{ij} X^{(r+3)\, a} \right)
\eeq
whence the $\cO(1)$ term in (\ref{Pot}) reduces to the standard
expression
\beq
f_{abe} f_{cde} \phi^{a\,ik} \phi^b{}_{jk} \phi^{c\,jl} \phi^d{}_{il}
 \, \propto \, {\rm Tr} \left([X^m , X^n]\right)^2\, .
\eeq
Conveniently, the conformal limit also disposes of the term
$\propto C^{ij} C_{ij}$ which makes the potential unbounded
from below. Again it is easy to see that the unboundedness of
the potential would persist with the `wrong' choice of gauge group.

In an analogous fashion we can show that in the Yukawa-like terms
in the Lagrangian only the term containing $C_{ab}{}^{ij}$ survives
the limit, and we end up with
\beq
-\frac{2i}{\ka} C_{ab}{}^{ij}\bar{\la}^a_i\la^b_j =
- {2i}f_{abc} \phi^{c\,ij}\bar{\la}^a_i\la^b_j + \cO(\ka)
\eeq
As before it follows that
\beq
f_{abc} \phi^{c\,ij}\bar{\la}^a_i\la^b_j + h.c.
\; \propto \; {\rm Tr} \, \bar{\la} \Gamma^m [ X^m, \la ]
\eeq
(where $\Gamma^m$ are the $SO(6)$ $\Gamma$-matrices), and similarly for
the remaining terms in the Lagrangian.

\section{Metamorphosis of the CW mechanism?}

As we already pointed out, the example of $N=4$ super Yang Mills falls
short of what we want in several respects (apart from anyway not being a
realistic model of particle physics). First, the flat space theory is
exactly conformally invariant: its $\beta$-functions vanishes to
all orders and there is no conformal anomaly (however, there are
non-trivial anomalous dimensions for composite operators). Secondly,
radiative breaking of symmetries cannot occur as a consequence of unbroken
supersymmetry. Finally, the non-conformal $N=4$ supergravity into
which this theory is embedded is not expected to be UV finite; hence,
the UV divergent gravitational corrections will destroy whatever
finiteness or conformal properties the flat space theory possesses.

Nevertheless there are some general conclusions that can be drawn from the
above construction and that may apply to other and hopefully more realistic
models.\footnote{See also \cite{OW} for an early attempt to construct
  a realistic model with broken supersymmetry by taking the
  $\ka\rightarrow 0$ limit and imposing restrictions similar to
  the ones discussed here.}
To obtain a classically conformal theory as a flat space limit from
a non-conformal UV completion of the SM, we must ensure that negative
powers of $\ka$ are absent before taking the limit $\ka\rightarrow 0$,
thereby also excluding a `bare' cosmological term $\propto \cO(\kappa^{-4})$.
As it happens, the $N=4$ theories discussed above do allow for a trivial
stationary point that meets all these requirements, but more interesting
extrema with non-vanishing expectation values of the scalar fields with
these properties appear not to exist for (\ref{GSO(n)}) or other
(non-compact) gauge groups \cite{BKS}. In addition, we would require the
stationary points to break supersymmetry completely in order to allow for
radiative breaking of symmetries. This is not an easy task to accomplish:
for instance, while the potential of $N=8$ supergravity (presumably the
only quantum field theoretic extension of Einstein's theory that may be
UV finite) does admit non-trivial stationary points \cite{Warner}, all
of these come with a non-vanishing cosmological constant
$\propto \cO(\kappa^{-4})$, and the ones that do break supersymmetry
are all unstable. It is therefore clear that `something extra' beyond
quantum field theory is required to avoid this {\em impasse}, and that
we must invoke an as yet unknown mechanism operating at the Planck scale
to make this idea work.

Independently of what the UV completion of the SM is, let us therefore
restate our main hypothesis: apart from the explicit $\cO(\ka)$ terms (which
can be neglected in the $\kappa\rightarrow 0$ limit) the UV completion
of the SM gives rise to finite logarithmic quantum corrections (depending
on $\log(\kappa\phi)$) which induce conformal symmetry breaking. These
terms would constitute an `observable' signature of quantum gravity,
in the sense that electroweak symmetry breaking might be entirely due to
finite (and in principle computable) quantum gravitational effects.
This immediately raises the question how such anomalous terms might
appear in the low energy effective action of a theory which is expected
to be UV finite. Conventional wisdom would suggest that there is no
need to regulate UV divergences in such a theory, hence there cannot
exist anomalies in the usual sense. The CW breaking would thus have to
come from {\em finite} logarithmic corrections induced by quantum gravity,
in accordance with the anomalous Ward identity (for $\kappa\rightarrow 0$)
\beq\label{Ward}
T^\mu{}_\mu = \beta\big(\hat\lambda(\kappa\phi)\big)\phi^4 +
   Z\big(\hat\lambda(\kappa\phi)\big)\partial_\mu\phi\partial^\mu\phi
\eeq
where the gravitational coupling $\kappa$ replaces the dimensionful
scale that must be introduced in quantum field theory in order to
parametrize the CW potential, and where the function $Z$ (which starts
only at higher orders in the couplings) and the effective running coupling
$\hat\lambda$  depend only logarithmically on their argument. While the
effective potential must obey a renormalization group equation to account
for the fact that this scale can be chosen arbitrarily, the gravitational
coupling $\kappa$ is a {\em fixed} parameter. The effective potential at
low energies can then be determined in principle by `integrating' the
anomalous Ward identity. We note that the potential importance of
(\ref{Ward}) for the solution of the hierarchy problem was already
emphasized in \cite{Bardeen}.

The existence of anomalous logarithmic terms in a UV finite theory is
also suggested by fascinating recent advances in the computation of
$n$-point amplitudes in $N=4$ super Yang Mills theory \cite{BDS,DHKS1,DHKS2}.
These have exposed (amongst other things) a new `dual' conformal
symmetry subtly intertwining IR and UV domains~\cite{AM,DHKS3,DHP}
as well as subtle `anomalous' effects in $n$-point correlators via
so-called cusp anomalies in the dual Wilson loops. Despite the exact
conformal invariance one obtains a non-vanishing result (conjectured
to be exact to all orders)
\beq
\Gamma_4(p_1,p_2,p_3,p_4)\propto \Gamma_{cusp}(g)
  \log^2 \left(\frac{p_1\!\cdot \! p_2}{p_2\!\cdot\! p_3}\right)
   \delta^{(4)}(p_1 + p_2 + p_3 + p_4)
\eeq
for the 4-point amplitude with $p_1^2=\cdots = p_4^2=0$ (this expression
is cyclically invariant because $p_1\!\cdot\! p_2 = p_3\!\cdot\! p_4$
and $p_2\!\cdot\! p_3= p_4\!\cdot\! p_1$). This result
is obtained by integrating
the anomalous Ward identity for the generator of conformal boosts
\cite{DHKS1,DHKS2}. Because the coupling $g$ in $N=4$ super Yang Mills
theory does not run, there can be no dynamically generated scale as in QCD,
and therefore the anomaly cannot appear in the effective potential (which
vanishes) but only in momentum dependent terms, that is, the non-local part
of the effective action. By contrast, gravity does possess a dimensionful
scale, so we can form the dimensionless quantity $\kappa\phi$, and anomalous
contributions in principle can show up in the {\em local} part of the
effective action. Let us also remark that it is much easier to arrange
for hierarchical expectation values if there are only logarithmic
corrections to the potential. This expectation is also supported by the
numerical finding that the minima of the CW effective potential \cite{MN1,MN4}
tend to be very shallow.

The main conjecture put forward in this paper can therefore be summarized
as follows: the hierarchy problem can conceivably be solved via `anomalous'
logarithmic quantum corrections in a UV finite theory of quantum gravity,
if the latter admits a flat space limit which is classically conformally
invariant. The mass spectrum and pattern of couplings observed in elementary
particle physics could then have their origin in quantum gravity.

\vspace{4.0mm}
\noindent
{\bf Acknowledgments:} We are grateful to E.~Bergshoeff, J.~Drummond and
E.~Sezgin for discussions and correspondence. K.A.M was partially
supported by the EU grant MRTN-CT-2006-035863 and the Polish grant
N202 081 32/1844, and thanks the Albert-Einstein-Institut for
hospitality and financial support.

\end{document}